# Uncertainty Quantification of the Fresh-Saltwater Interface from Time-Domain Electromagnetic Data

Arsalan Ahmed[1], Thomas Hermans[1,*], David Dudal[2], and Wouter Deleersnyder[1,2,3]

1. Department of Geology, Ghent University, Krijgslaan 297, 9000 Gent, Belgium;
2. Department of Physics, KU Leuven KULAK, Etienne Sabbelaan 53, 8500 Kortrijk, Belgium;
3. Geophysical Inversion Facility, University of British Columbia, Vancouver, Canada;

* Correspondence: thomas.hermans@ugent.be (T.H)

## Abstract

Geophysical methods offer a cost-effective way to characterize the subsurface for hydrogeological projects, but they rely on processing the data through an inverse problem. Traditionally, the inverse problem is solved with deterministic methods, which pose challenges due to the non-uniqueness of the solution. Stochastic approaches provide a more comprehensive view of the solution by quantifying the uncertainty, albeit requiring more computational resources. Bayesian Evidential Learning (BEL) circumvents the stochastic inversion process by using a machine learning approach, providing an approximation of the posterior distribution at a lower cost. Still, as with most Monte Carlo sampling techniques, the efficiency of the approach and its computational costs highly depend on the number of inversion parameters. In this contribution, we demonstrate that by incorporating available prior knowledge into the parameterization of the inverse problem, the number of unknowns can be effectively reduced, hence reducing the computational burden. We exemplify this on time-domain electromagnetic data to identify fresh and saltwater interfaces in the Flemish (Belgium) coastal aquifer. Most traditional blocky and smooth deterministic inversion techniques struggle with properly recovering this transition zone, representing it as either too sharp or too gradual. To address this problem, we parameterize the transition zone with only two parameters, namely its depth and its thickness, assuming a linear transition from fresh to saltwater. This parameterization retains the compactness of (blocky) parametric inversion, while allowing for both sharp and gradual interfaces as with voxel-based inversion. To learn more about the reliability of the recovered transition zone, we invert these parameters stochastically using Bayesian Evidential Learning with Thresholding (BEL1D-T). Results show that this parameterization in BEL1D-T effectively outlines the uncertainty range for both synthetic and field data. The transition zone remains largely uncertain due to the limitations of the survey design and the non-uniqueness of the inverse problem. Using our probabilistic method proves effective, without the significant computational costs associated with traditional probabilistic approaches.

# 1 Introduction

Sustainably managing groundwater resources in coastal areas requires hydrogeologists to determine the position of the fresh-saltwater interface (FSI) and monitor its temporal variations (Kim et al. 2006; Werner et al. 2013; Bear et al. 1999). Knowing the FSI in coastal aquifers is vital to control the accessibility to freshwater and the environmental well-being of coastal ecosystems (Goebel et al. 2021; Carrera et al. 2010).

Complex subsurface features and dynamic subsurface processes make it difficult to accurately characterize the FSI, which commonly differs from the wedge shape described in textbooks and only encountered in simpler coastal geologies (Goebel et al. 2021). The position and shape of the FSI in coastal aquifers depend on multiple factors (Kim et al. 2006), including the density difference between freshwater and saltwater (Nguyen et al. 2009), freshwater recharge rates, groundwater extraction (Guo and Jiao 2007), the hydrogeological evolution of aquifers related to seawater levels (Dieu et al. 2022), aquifer heterogeneity (Cong-Thi et al. 2024), and structure (Paepen et al. 2025), leading to complex shapes of the FSI (Goebel et al. 2019). The FSI can be a sharp boundary or a diffuse mixing zone where salinity changes gradually through a transition zone (e.g., Kim et al. 2006; Bear et al. 1999). Moreover, some irregular configurations and salinity reversals can be seen in complex multi-aquifer systems (Yechieli et al. 2001) or in subterrain estuaries, (e.g., Paepen et al. 2020). The transition zone between freshwater and saltwater thus becomes quickly untraceable by conventional borehole observations because of its lateral and vertical variability (Carrera et al. 2010).

Geophysical methods offer a cost-effective solution to study the subsurface. Multiple studies showed that the time-domain electromagnetic method (TDEM) is effective for seawater intrusion and FSI mapping in coastal aquifers (Kafri et al. 2007; Goldman et al. 1991; Yechieli et al. 2001; Kafri and Goldman 2005, Goebel et al., 2019; Delsman et al., 2018; King et al., 2018), due to its sensitivity to conductive bodies. The large-scale mapping capability of airborne (e.g., Goebel et al. 2019) or towed systems (Auken et al. 2019) provides valuable insights into the FSI in complex coastal plains.

However, TDEM data must be further processed by solving an inverse problem to provide information on the FSI. The problem is ill-posed, and the solution is non-unique, meaning that multiple possible models can fit the observed data. Deterministic inversion typically requires additional constraints imposed via regularization methods to provide a solution. For example, blocky inversion yields models with distinct boundaries between layers. Alternatively, smoothness-constraint regularization produces inversion models with gradual transitions (Constable et al., 2015). The inversion is ften based on the discretization of the subsurface into a large number of relatively thin layers. The inverted models often then exhibit excessive sharpness or smoothness (Deleersnyder et al. 2021).

While the traditional deterministic methods produce a single solution and have proven robust, they fail to assess the uncertainty. Stochastic inversion methods provide an alternative for exploring the parameter spaces and assessing uncertainty (Sambridge and Mosegaard 2002; Trainor-Guitton and Hoversten 2011; Ball et al. 2020). However, their computational cost quickly grows with the number of parameters, while their results remain dependent on the selection of the prior distribution of model parameters (Ahmed et al. 2024; Aigner et al., 2025).

When one is interested in a specific target, such as a particular interface, it can be advantageous to integrate this interface into the parameterization itself. This can be done through a hierarchical approach, where the interface is first defined as a boundary between different zones, and then variability within each zone is solved separately. De Pasquale et al. (2019) used such an approach to invert for the depth of the bedrock from ERT data, but the interface itself was restricted to a predefined grid. Alternatively, the interface itself can be parameterized so that the inversion focuses on retrieving the parameters that define this interface. For example, Goebel et al. (2021) use a parametric inversion to invert the saltwater wedge from electrical resistivity data. The interface was represented as a 5$^{th}$-order polynomial, with fixed resistivity above and below the threshold and was translated into a predefined regular grid. If these approaches can effectively represent the interface itself, they ignore the gradual transition characteristic of the FSI.

Indeed, in coastal areas, the upper or phreatic aquifer salinity levels tend to rise progressively from a shallow freshwater lens with almost constant salinity (Vandenbohede et al., 2015), originating from rainwater infiltration, to a denser saltwater layer of relatively constant salinity. Considering the characteristics of this gradual transition from fresh to saltwater, we introduce a new parameterization of the FSI, based on its depth and rate of conductivity increase, allowing a compact representation of the model parameters. We combine it with a stochastic approach to quantify the uncertainty related to the FSI from TDEM data, namely Bayesian Evidential Learning with thresholding (BEL1D-T) (Ahmed et al. 2024; Aigner et al. 2025). BEL1D-T is based on learning a direct relationship between data and model parameters in a low-dimensional space. Previous applications on geophysical inversions have focused on inversion of 1D layered models (Michel et al. 2020; 2022a; Ahmed et al. 2024; Aigner et al. 2025). We investigate the influence of FSI characteristics (sharpness and depth) as well as the presence of an aquitard at a known depth on the uncertainty. We illustrate the methodology with synthetic data and some field surveys caaried out in the Belgian coastal area.

In the next section, we outline our dedicated parameterization, the TDEM forward modeling, and the stochastic inversion approach BEL1D-T. Next, we present the validity of our proposed approach for imaging the FSI with synthetic cases covering various depth and sharpness of the FSI before discussing the results from a real dataset. We finish the paper with conclusions and a future outlook.

## 2. Methodology

To determine the uncertainty of the FSI from TDEM data, we propose a dedicated parameterization in Section 2.1 of the electrical conductivity of the subsurface, including a transition zone between fresh and saltwater and the presence of a confining clay layer at depth. Although this is based on the expected distribution of the Belgian coastal zone, this parameterization can be extended to other contexts. For this parameterization, a prior distribution is constructed by examining all available prior knowledge. The posterior distribution is obtained via the BEL1D-T method, outlined in Section 2.3, which efficiently generates an ensemble of posterior models (Ahmed et al. 2024). Without loss of generality, we illustrate the approach for TDEM data acquired with the ABEM WalkTEM system.

### 2.1 Parameterization

The objective of our parameterization is to represent the electrical conductivity distribution with a limited number of parameters, knowing the characteristic transition from fresh to saltwater with an FSI in between. The parameterization (Figure 1A) consists of 6 parameters: the thickness ($t_1$) and electrical conductivity ($\sigma_1$) of the freshwater layer, the thickness of the transition zone ($t_2$), the thickness ($t_3$) and electrical conductivity ($\sigma_3$) of the saltwater layer, and the conductivity of the half-space ($\sigma_4$). The parameter $t_1$ also corresponds to the depth of the FSI, while the sum of the three thicknesses $t_1 + t_2 + t_3$ corresponds to the depth $d_4$ of the half-space. This parameterization is effectively a dimensionality reduction, compared to traditional voxel-based inversion which would typically use a few tens of parameters to solve the 1D problem with regularization (Ball et al., 2020). It is therefore particularly beneficial for computationally demanding stochastic methods.

The parameterization was inspired by the Belgian coastal aquifer context (Vandenbohede and Lebbe 2011). In particular, the effectiveness of the parameterization to capture change in conductivity was verified using electromagnetic log measurements collected in boreholes (Figure 1b). Within this context, the half-space corresponds to thick clay layers, which constitute the confining layer of the Quaternary aquifer. Boreholes are typically drilled down to this clay confining layer. The proposed parameterization ignores variations occurring within each layer, as well as the deviation from a linear increase in conductivity. However, these variations are below the resolution of surface geophysical methods and are also missed by regularization-based inversion (e.g., Deleersnyder et al., 2021). The parameterization also ignores the presence of a more resistive vadose zone. This is justified by the presence of a shallow water layer in the study area (less than 2m). The parametrization can be adapted in other contexts if needed. The addition of a transition zone or a layer only requires one or 2 extra parameters respectively.

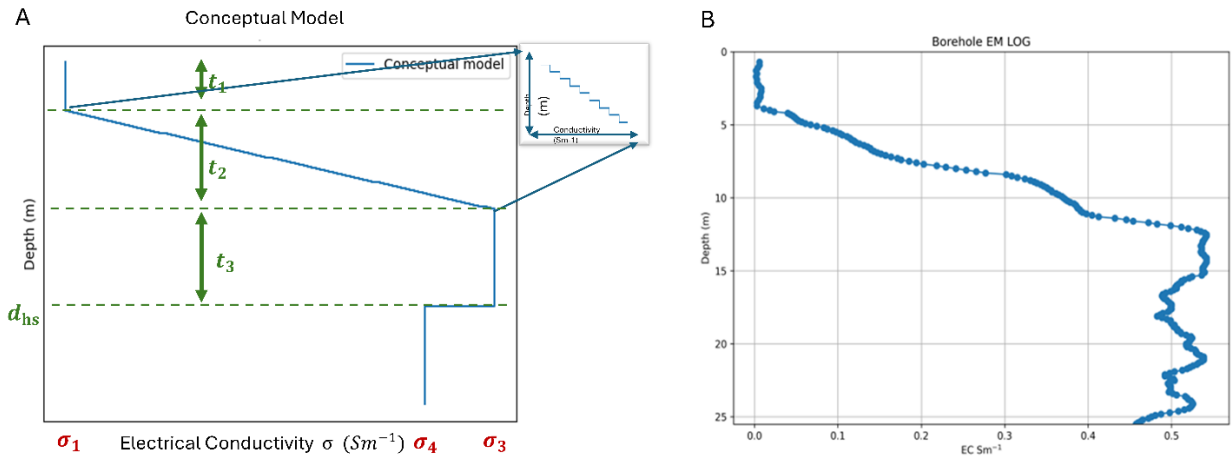

*Figure 01: A. Conceptual parameterization of the conductivity model using 6 parameters. B. Electromagnetic log showing the actual electrical conductivity distribution in the study area.*

## 2.2 Time domain electromagnetic method and forward modeling

Time-domain electromagnetic (TDEM) is based on the principle of electromagnetic induction (EMI). A quasi-instantaneous switch-off of a magnetic dipole (the primary field) generates a (so-called secondary) magnetic field, due to induced currents in the (slightly) conductive subsurface. That secondary magnetic field depends on the electrical conductivity distribution of the subsurface, which we aim to image by processing the observed decaying secondary magnetic field. This requires an efficient forward model, which describes the subsurface response to such a magnetic dipole, given the parameter distribution of the subsurface and the setup of the measurement instrument. We resort to the semi-analytical 1D solution developed by Hunziker and Slob (2015), which assumes horizontal and isotropic layers. Although this might cause modeling errors (Deleersnyder et al. 2024), it is considered sufficient for this application, as only 1D subsurface models are considered. We use the empymod implementation of the forward model (Werthmüller 2017), which is an open-source Python package that efficiently produces electromagnetic responses. Note that the transition zone must be further discretized into a set of smaller layers, to be able to be evaluated by the forward model, see Figure 1A.

## 2.3    Bayesian Evidential Learning 1D with thresholding

Bayesian Evidential Learning in 1D with thresholding (BEL1D-T) is used to solve the inverse problem stochastically. A statistical relationship obtained in a low dimensional space that connects observed data to the target parameters is derived from a set of prior models sampled from the prior distribution. The inverse problem is solved stochastically in the low dimensional space instead of using regularization to handle the ill-posedness of the inverse problem.

In this paper, we learn a direct relation between the TDEM decay curves and the six parameters following the parameterization of the FSI. Empymod is used to generate the TDEM decay curve for each sample from the prior distribution. To reduce the number of needed training samples, BEL1D learns the statistical relationship in a low-dimensional space by performing PCA on the data space, including propagation of the expected noise level (Hermans et al., 2016). The dedicated parameterization guarantees the low dimensionality of the model space. We use Canonical Correlation Analysis (CCA) to learn a linear statistical relationship between model parameters and PCA data coefficients. Although CCA is based on linear combinations, it can handle slight non-linearities in the data-target relationship. It is however most efficient for a low number of dimensions, which is guaranteed by the use of PCA and the proposed parameterization. Note that, if CCA would fail, other non-linear methods can be used, such as Probabilistic Neural Networks (Zhang et al., 2026). BEL1D can then generate the posterior distribution for any observed data consistent with the prior distribution.

Because BEL1D does not use any forward simulations during inference of the posterior distribution, it can lead to an overestimation of the uncertainty when prior uncertainty is large (Michel et al., 2020). To overcome this problem, Michel et al. (2023) suggested using Iterative Prior Resampling (IPR) to reach similar results to those of a Markov chain Monte Carlo (MCMC) solution, albeit at the cost of additional forward model computations. Recently, Ahmed et al. (2024) have demonstrated that using a filter based on threshold on the relative root-mean-squared-error (rRMSE) of BEL1D posterior models was an efficient alternative to quickly approximate the posterior distribution (BEL1D-T). A stricter threshold results in a lower posterior range. Aigner et al. (2025) suggested to set the threshold larger than the estimated noise level to account for the typical underestimation of noise from stacking in TDEM data, the uncorrelated nature of noise in TDEM data and still accept some models with a lower likelihood. Doing so, BEL1D-T solution can be considered as fast approximation of the posterior distribution. Another advantage is that the same training set can be reused for multiple soundings within the study area as long as the prior remains valid (Aigner et al. 2025). More details on BEL1D and BEL1D-T can be found in Michel et al. (2020) and Ahmed et al. (2024) respectively.

## 3. Results

We first test the methodology on synthetic cases to evaluate how different parameter combinations affect the uncertainty of the recovered FSI. Section 3.2 applies our proposed methodology to a field data case at the Molenkreek, northwest Belgium. The synthetic cases are set to mimic the field data case which uses the ABEM WalkTEM 2 set-up (ABEM instruments, 2012). The transmitter coil consists of a single 20 m x 20 m square loop, while the receiver has a 0.5 m$^2$ rigid coil design placed in the center of the transmitter. Note that the WalkTEM is a dual moment system. The high moment generates a stronger magnetic field, which results in a larger signal-to-noise ratio at larger depths or late times. The low moment has a shorter turn-off time, allowing the use of earlier time gates and imaging shallower structures (Chongo et al. 2015). For the low moment, the current was 1 *A* and the first data point was recorded at 12,79 µs. A current of 12 *A* was used for the high moment, with a first available data point at 15,10 µs.

## 3.1 Synthetic cases

In the synthetic case, we examine 3 cases with increasing complexity and realism. For each case, we investigate several configurations of the FSI, namely various depths and sharpness.

1. **Case 1: Unknown thicknesses, known conductivities.** The electrical conductivity of the fresh, salt, and underlying clay layers is assumed to be known and stays fixed during the inversion. Only the thicknesses are unknown within their prior range.

2. **Case 2: Unknown model parameters**. All thicknesses and conductivities from the parameterization are unknown.

3. **Case 3: Prior knowledge on the depth of the half-space.** This problem is similar to Case 2, except for an extra constraint for the depth of the half-space. This is the depth at which the saltwater rests on the underlying clay layer, which is often known from existing geological models, and hence, according to Bayes' theory, should be incorporated into the stochastic inversion.

For each case, 3% multiplicative Gaussian noise is added to the synthetic data. In the BEL1D-T, the rRMSE threshold is set to 0.135, which removes all models with an rRMSE larger than 15% (see Ahmed et al., 2024). The effect of the threshold is investigated for the field case. We use 5000 samples from the prior distributions for training. Results are presented using diagonal plots showing uni- and bivariate posterior distribution, with depth-conductivity profiles and with boxplots to more easily compare the results.

### 3.1.1 Case 1: Unknown Thicknesses, known conductivities

In this case, only the thickness of each layer, including the transition zone, is a free parameter in the inversion, while all conductivities are considered known and fixed. We evaluate the solution and uncertainty for four realistic FSI scenarios. We consider two different depths at which the FSI starts: shallow at 5 m or deep at 15 m. Additionally, two different transition geometries are considered: a sharp transition with a FSI thickness of 2.5 m and a gradual transition of 10 m thickness. The true model for each scenario, together with the prior parameter ranges, is tabulated below in Table 1 and correspond to realistic situations encountered on the field.

*Table 1: The uniform prior distribution and the true value for model parameters used for the inversion of Case 1. The $t_i$ and $\sigma_i$ are the thickness and the electrical conductivity of each layer i, respectively.*

| FSI Scenario | Parameters | | | | | |
| --- | --- | --- | --- | --- | --- | --- |
| | $t_i$ and $\sigma_i$ are the thickness and the electrical conductivity of each layer $i$, respectively. | | | | | |
| Shallow/sharp | $t_1 \in [1, 20]$ True $t_1$: 5 | $t_2 \in [0.5, 15]$ True $t_2$: 2.5 | $t_3 \in [1, 36]$ True $t_3$: 25 | / True $\sigma_1$: 0.05 | / True $\sigma_3$: 0.6 | / True $\sigma_4$: 0.4 |
| Shallow/Gradual | $t_1 \in [1, 20]$ True $t_1$: 5 | $t_2 \in [0.5, 15]$ True $t_2$: 10 | $t_3 \in [1, 36]$ True $t_3$: 25 | / True $\sigma_1$: 0.05 | / True $\sigma_3$: 0.6 | / True $\sigma_4$: 0.4 |
| Deep/Sharp | $t_1 \in [1, 20]$ True $t_1$: 15 | $t_2 \in [0.5, 15]$ True $t_2$: 2.5 | $t_3 \in [1, 36]$ True $t_3$: 25 | / True $\sigma_1$: 0.05 | / True $\sigma_3$: 0.6 | / True $\sigma_4$: 0.4 |
| Deep/Gradual | $t_1 \in [1, 20]$ True $t_1$: 15 | $t_2 \in [0.5, 15]$ True $t_2$: 10 | $t_3 \in [1, 36]$ True $t_3$: 25 | / True $\sigma_1$: 0.05 | / True $\sigma_3$: 0.6 | / True $\sigma_4$: 0.4 |

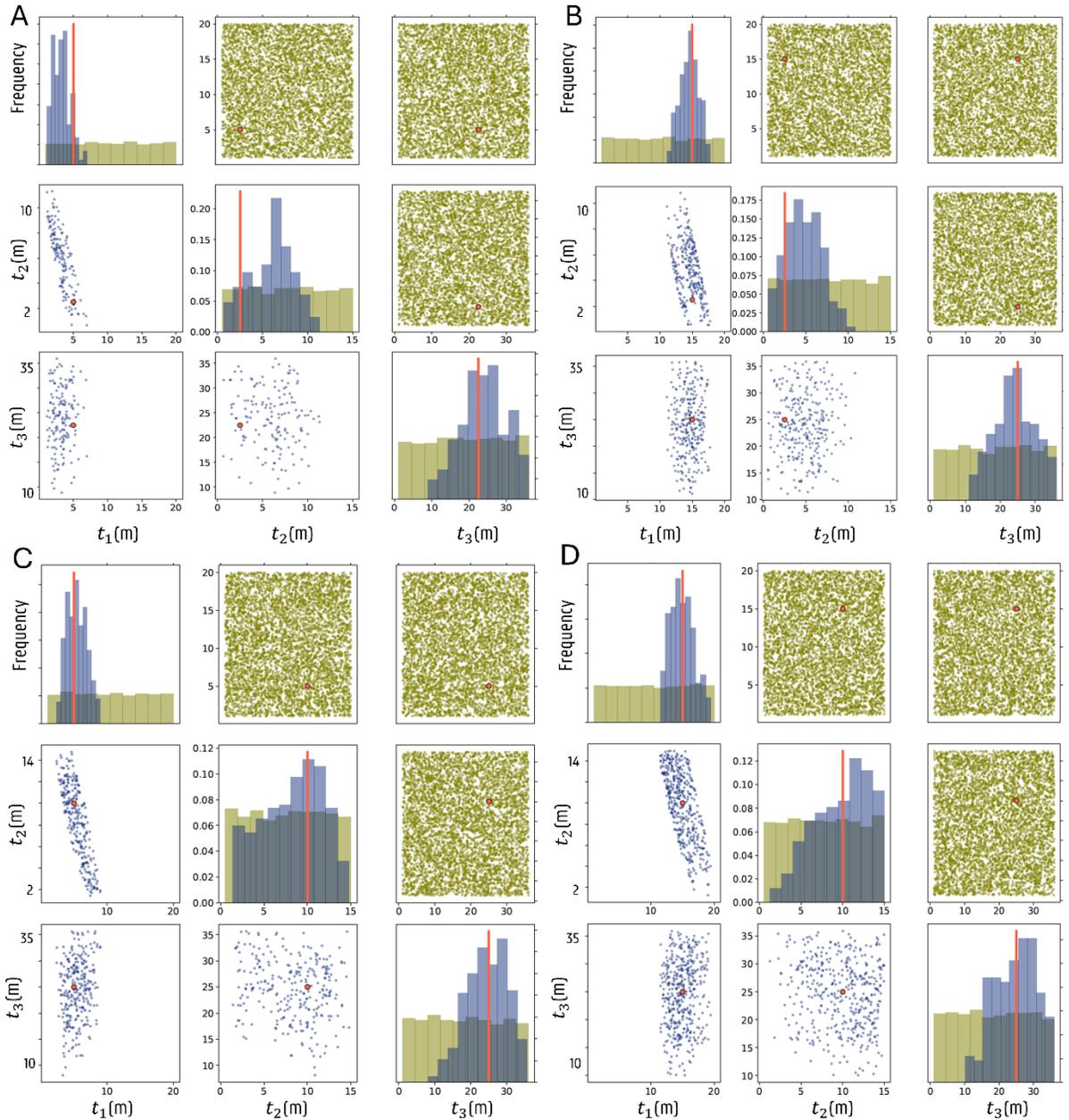

*Figure 2. Visualization of the posterior model space for shallow/sharp (A), shallow/gradual (B), deep/sharp (C) and deep/gradual (D) transitions. The mustard-colored scatter plot represents the samples from the uniform prior distribution, while the blue scatter plot and histograms illustrate the posterior model distribution. The red dots and lines indicate the true model.*

The results show the difficulty of accurately recovering the FSI from the TDEM data (Figures 2 to 4). The depth of the interface is properly retrieved, but with an uncertainty range of a few m. The posterior is slightly biased toward lower depths for the shallow/sharp interface as clearly visible on Figures 3 and 4, and to some extent for the deep/sharp interface. This difficulty in imaging the shallow interface is expected due to the electronic limitations of most TDEM instruments. It is impossible to switch off the primary magnetic field instantaneously; hence, the very early time gates, sensitive to the uppermost part of the subsurface, cannot be accurately recorded. Therefore, the inversion struggles to precisely capture transitions that are 5 m deep or less precisely. The range of uncertainty is not strongly affected by the depth of the FSI itself, as deeper transitions show a similar uncertainty range as the shallow ones. The thickness of the transition zone $t_2$ has a considerable posterior uncertainty for all transition combinations. The deep/sharp and shallow/sharp transitions display an overestimation of the thickness of the transition zone, with a bias toward higher values, clearly highlighting that the sensitivity of the TDEM to sharp interfaces is limited (Figure 4). This confirms existing research, where an increased uncertainty in the region right on top of the saltwater lens is attributed to the inherent non-uniqueness of the inverse problem (King, 2022; Deleersnyder et al., 2024b).

Figure 2 also clearly identifies a correlation in the posterior distribution between the depth and the thickness of the FSI ($t_1$ and $t_2$). When the former increases, the latter is reduced. It therefore seems that the data set is mostly sensitive to the depth of the saline layer, corresponding to $t_1 + t_2$, rather than the FSI itself, which is coherent with the sensitivity of TDEM increasing in conductive layers.

The thickness of the saline layer is properly captured for the tested cases by the mean of the posterior distribution, but with a large uncertainty (Figure 4). Only very low thicknesses are properly identified as not realistic by the TDEM data. The limited contrast in conductivity with the underlying clay layer can explain this.

In practice, the depth of the saline interface is often defined based on a threshold value of electrical conductivity (e.g., Delsman et al., 2019) and not based on the thickness of the freshwater layer as proposed above. The choice of thresholding, generally based on some petrophysical assumption and a threshold value for salinity, can have a significant impact on the estimation of the uncertainty of the FSI depth (Deleersnyder et al. 2024). Figure 3 illustrates this with a minimum uncertainty range for the middle of the transition zone around a conductivity of 0.3 S/m compared to the top of the FSI at 0.05 S/m.

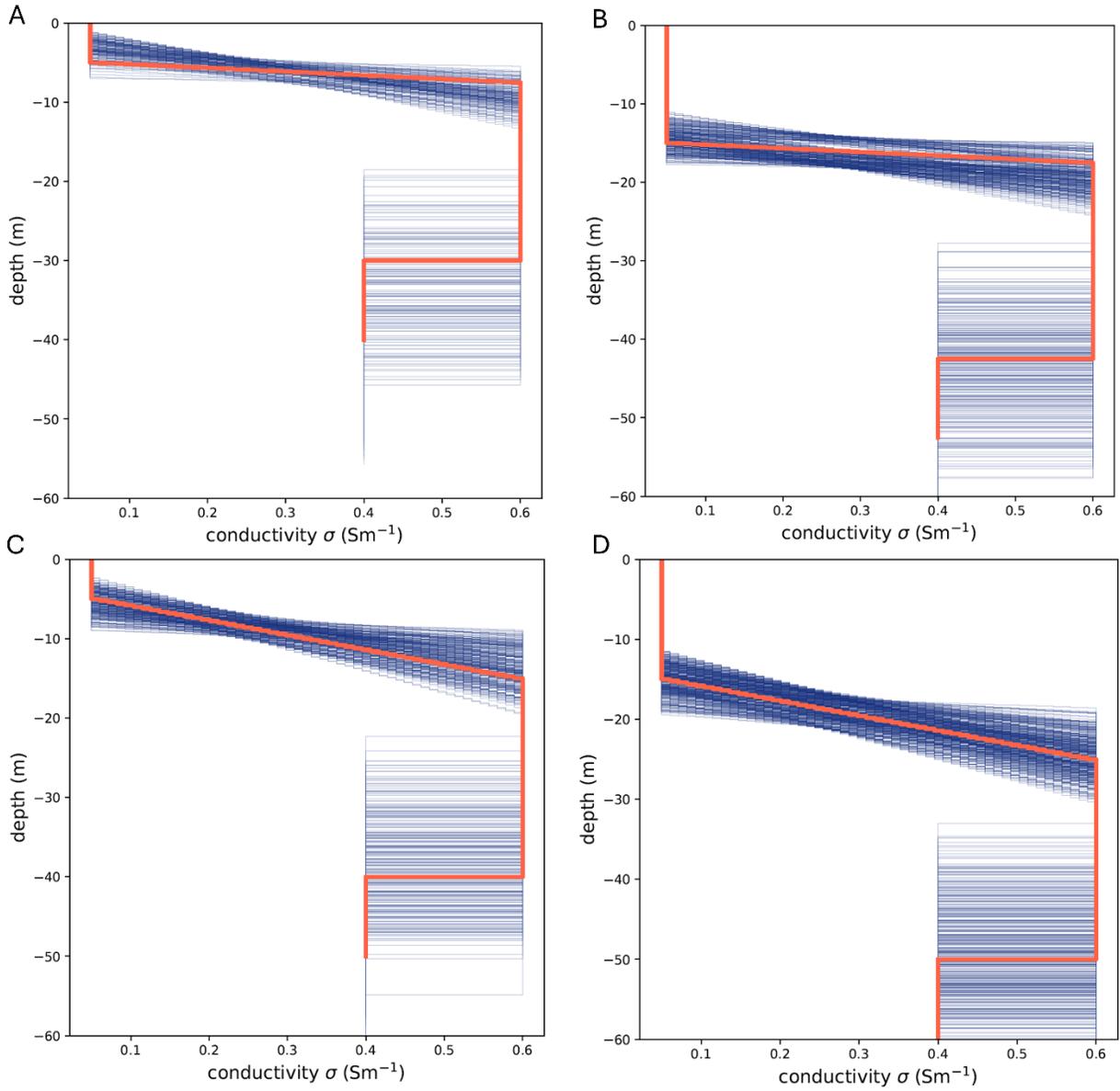

*Figure 3. Posterior depth (m) vs conductivity (Sm$^{-1}$) model of shallow/sharp (A), shallow/gradual (B), deep/sharp (C) and deep/gradual (D) transitions.*

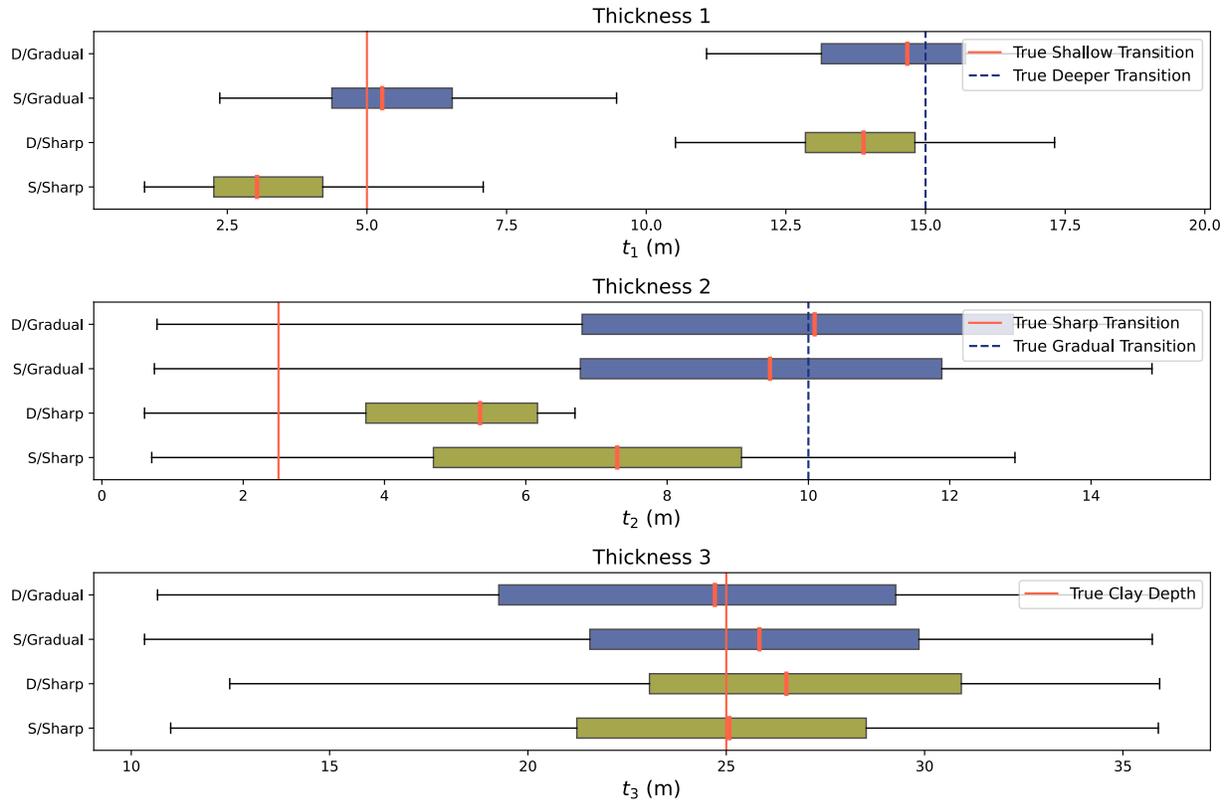

*Figure 4. Comparison of transition types across the thicknesses of layers 1, 2, and 3. The red and dotted blue lines indicate the true values.*

### 3.1.2. Case 2: Unknown model parameters

In this section, we investigate how the interface or transition zone uncertainty changes by releasing the constraint on the conductivities. In contrast to the previous case, the conductivities become parameters in the inversion. For the sake of conciseness, we limit the results to the deep interface; similar results were obtained for the shallow interface.

*Table 2 The uniform prior distribution and the true value for model parameters used for the inversion of Case 2. The $t_i$ and $\sigma_i$ are the thickness and the electrical conductivity of each layer i, respectively.*

| FSI Scenario | Parameters | | | | | |
|---|---|---|---|---|---|---|
| Deep/Sharp | $t_1 \in [1, 20]$ True $t_1$: 15 | $t_2 \in [0.5, 15]$ **True $t_2$: 2.5** | $t_3 \in [1, 36]$ True $t_3$: 25 | $\sigma_1 \in [0.01, 0.1]$ True $\sigma_1$: 0.05 | $\sigma_3 \in [0.25, 1]$ True $\sigma_3$: 0.6 | $\sigma_4 \in [0.3, 0.5]$ True $\sigma_4$: 0.4 |
| Deep/Gradual | $t_1 \in [1, 20]$ True $t_1$: 5 | $t_2 \in [0.5, 15]$ **True $t_2$: 10** | $t_3 \in [1, 36]$ True $t_3$: 25 | $\sigma_1 \in [0.01, 0.1]$ True $\sigma_1$: 0.05 | $\sigma_3 \in [0.25, 1]$ True $\sigma_3$: 0.6 | $\sigma_4 \in [0.3, 0.5]$ True $\sigma_4$: 0.4 |

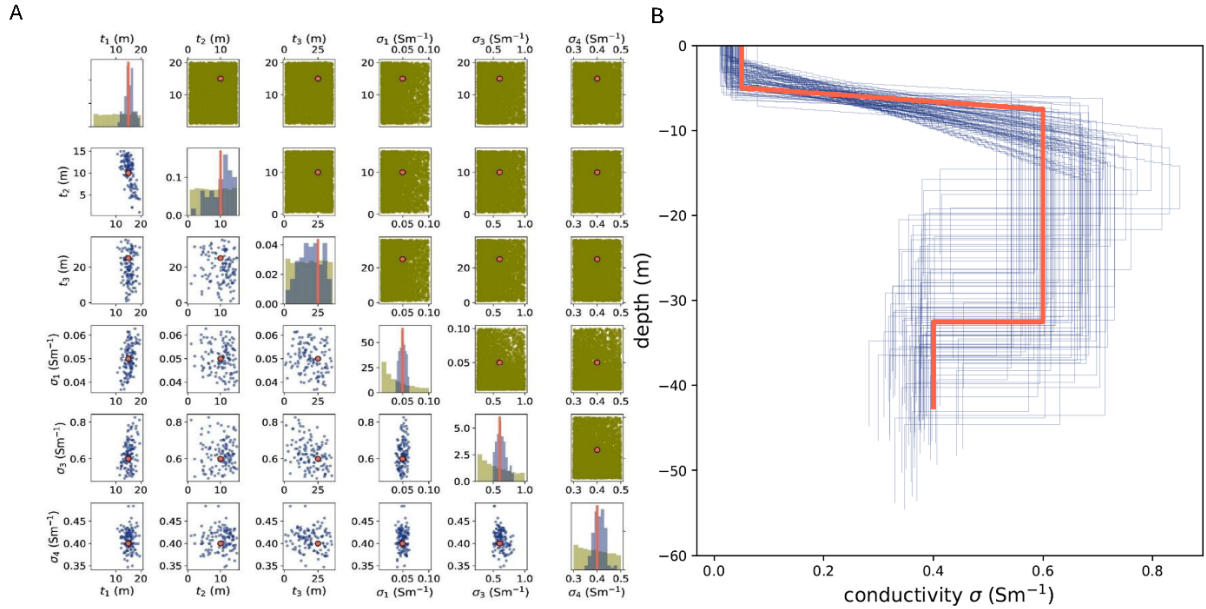

*Figure 05. Posterior space (A) and depth/conductivity model visualization (B) for the deep/gradual transition for case 2.*

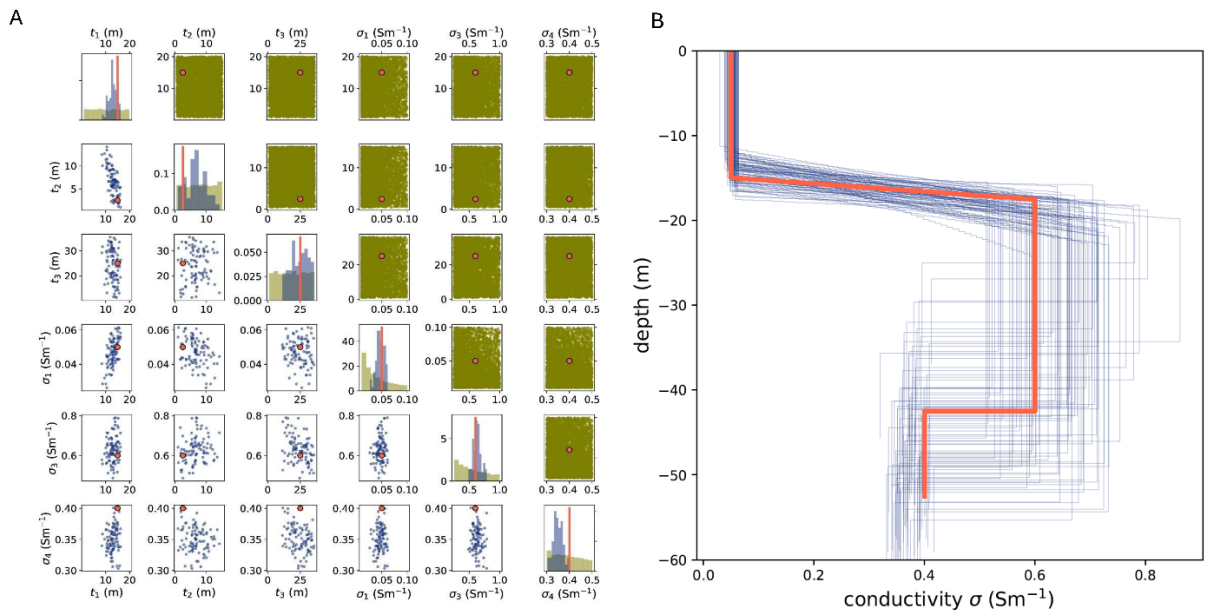

*Figure 6. Posterior space (A) and depth/conductivity model visualization (B) for the deep/sharp transition for case 2.*

The inversion results (Figures 5 and 6) demonstrate that the approach successfully determines the conductivities of the different layers with limited uncertainty. In contrast to the deep/gradual scenario, the deep/sharp scenario sees the conductivity of the half-spaced biased towards lower conductivity values. This observation could be explained by the fact that the amplitude of the TDEM data is increasing with conductivity. For the deep/sharp case, the half-space is reached at lower depth, leading to a lower signal recorded at later times and a lower signal-to-noise ratio. However, an opposite observation is observed for the shallow interfaces (Figure A2.1). For the shallow interfaces, the uncertainty on the first layer conductivity is larger than for deep interfaces, because the system configuration is less sensitive to shallow layers (Figure A2.3). A correlation remains between the thickness of the fresh layer and the transition zone (Figures 5 and 6) as observed in case 1. For the deep/sharp FSI, some correlation also exists between the thickness ($t_3$) and conductivity ($\sigma_3$) of the saline layer.

Figure 7 compares the posterior distribution of the thicknesses with (case 2) and without (case 1) uncertainty on the conductivity of the layers. The results indicate that the FSI depth can be determined with a similar precision and accuracy, independent of the uncertainty on the electrical conductivity. The biased estimation for the thickness of the FSI for the sharp interface remains similar. This indicates that the uncertainty of the FSI is not strongly affected by the conductivity value above and below the interface.

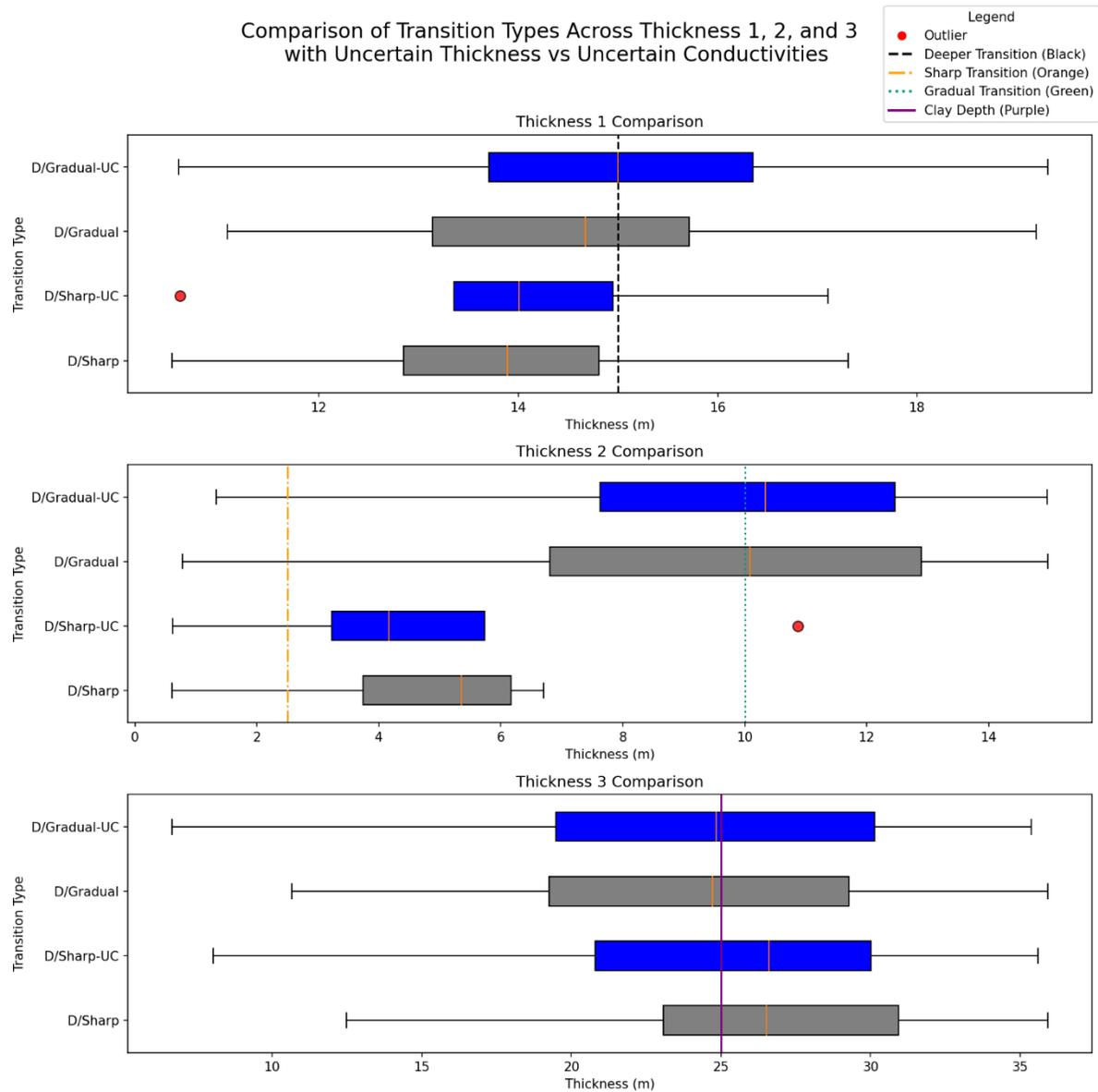

*Figure 7. Comparison of the posterior thicknesses of layers 1, 2, and 3 with and without uncertain conductivities (UC) for Deep/Sharp and Deep/Gradual transitions. The dotted black line indicates the depth of the transition zone. The orange and green lines represent the thicknesses of the transition layers, with the orange line denoting a sharp transition zone of 2.5 meters and the green line indicating a gradual zone of 10 meters. The purple line highlights the thickness of the third layer (saline layer).*

### 3.1.3 Case 3: Prior knowledge on the depth of the half-space

Figures 3, 5, and 6 show that the uncertainty on the depth of the half-space, corresponding to the some of thicknesses of the 3 layers, remains large after inversion. In practice, a good estimation of the depth of the first aquitard layer (clay), corresponding to the half-space in this case study, is often available from drilling logs. In this section, we use the depth of the half-space as a source of prior information to constrain the inversion. This requires a slight re-parameterization of the problem, where the depth of the half-space ($d_{hs}$) becomes an inversion parameter, while the thickness of the saltwater lens is obtained through $t_3 = d_{hs} - t_2 - t_1$.

*Table 3. The uniform prior distribution and the true value for model parameters used for the inversion of Case 3. The $t_i$ and $\sigma_i$ are the thickness and the electrical conductivity of each layer i, respectively, while $d_{hs}$ is the constrained depth of the half-space.*

| FSI Scenario | Parameters | | | | | |
|---|---|---|---|---|---|---|
| Deep/Sharp | $t_1 \in [1, 20]$ True $t_1$: 15 | $t_2 \in [0.5, 15]$ True $t_2$: 2.5 | $d_{hs} \in [22, 28]$ True $t_3$: 25 | $\sigma_1 \in [0.01, 0.1]$ True $\sigma_1$: 0.05 | $\sigma_3 \in [0.25, 1]$ True $\sigma_3$: 0.6 | $\sigma_4 \in [0.3, 0.5]$ True $\sigma_4$: 0.4 |
| Deep/Gradual | $t_1 \in [1, 20]$ True $t_1$: 5 | $t_2 \in [0.5, 15]$ True $t_2$: 10 | $d_{hs} \in [22, 28]$ True $t_3$: 25 | $\sigma_1 \in [0.01, 0.1]$ True $\sigma_1$: 0.05 | $\sigma_3 \in [0.25, 1]$ True $\sigma_3$: 0.6 | $\sigma_4 \in [0.3, 0.5]$ True $\sigma_4$: 0.4 |

Figure 8 compares the thicknesses of the 3 layers with (case 3) and without (case 2) constraint on $d_{hs}$. Globally, this constraint yields only limited changes for the first layer for both deep/gradual and deep/sharp FSI. The bias for the depth deep/sharp FSI is reduced when introducing a clay depth constraint. For the thickness of the transition zone, the clay depth constraint decreases the average thickness to a value closer to the true one, while also reducing the posterior range. For the deep/gradual FSI, the clay depth constraint induces a small bias in estimating the thickness of the transition zone towards lower values.

The most significant impact of the clay constraint is visible on the estimated thickness of layer 3. Thanks to the constraint, the uncertainty is largely reduced. While the posterior distribution without constraint (Figures 5 and 6) covered almost the whole prior range, adding the constraint reduced the uncertainty to a range of about 5 m. However, this decrease in uncertainty does not strongly impact the inversion of the FSI itself. Similar observations are made for shallow interfaces (appendix).

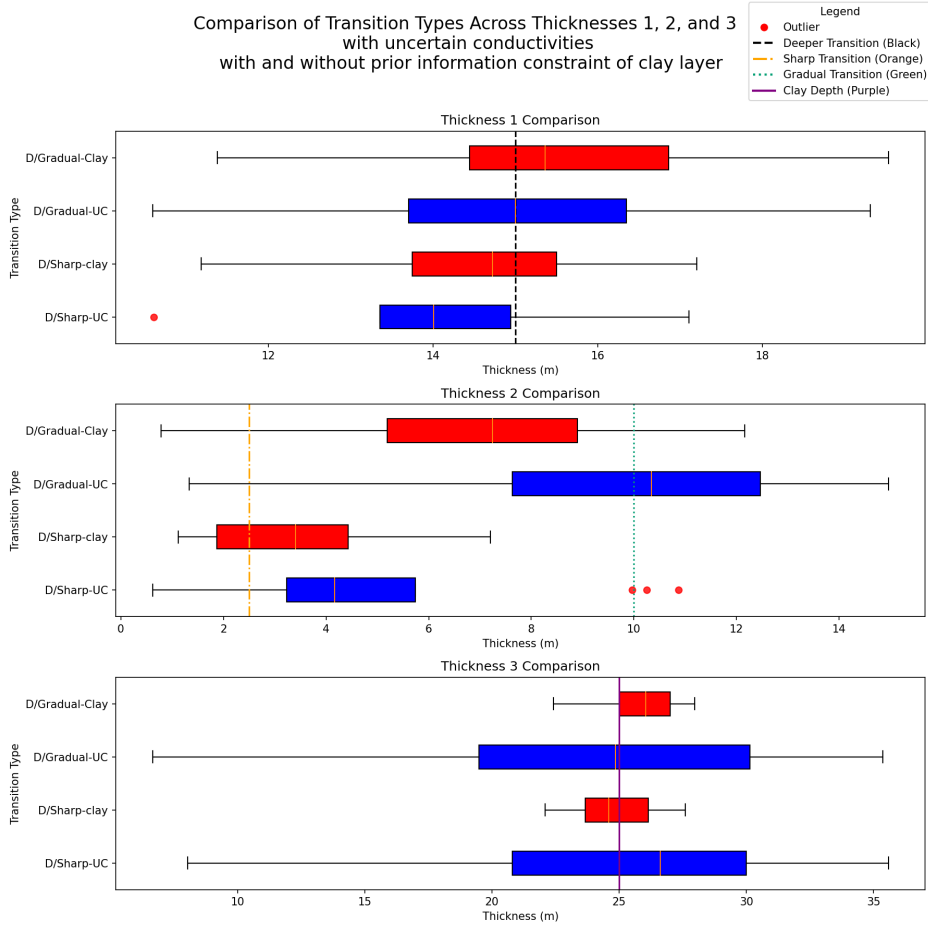

*Figure 8. Comparison of the posterior range for the thicknesses for Deep/Sharp and Deep/Gradual transition with and without clay depth constraint*

## 3.2 Field Case Study

The Molenkreek area is located in the Meetjesland, a polder area within the Belgian coastal plain (Blanchy et al. 2025). Ground-based TDEM data were acquired to compare the results with AEM data used for the salinity map of Belgium (Figure 9). The acquisition set-up was identical to the synthetic case. To estimate the measurement noise level, five stacks of data were recorded without any magnetic moment, and 50 stacks were recorded for each moment. For more details about this standard procedure, we refer to the ABEM WalkTEM manual.

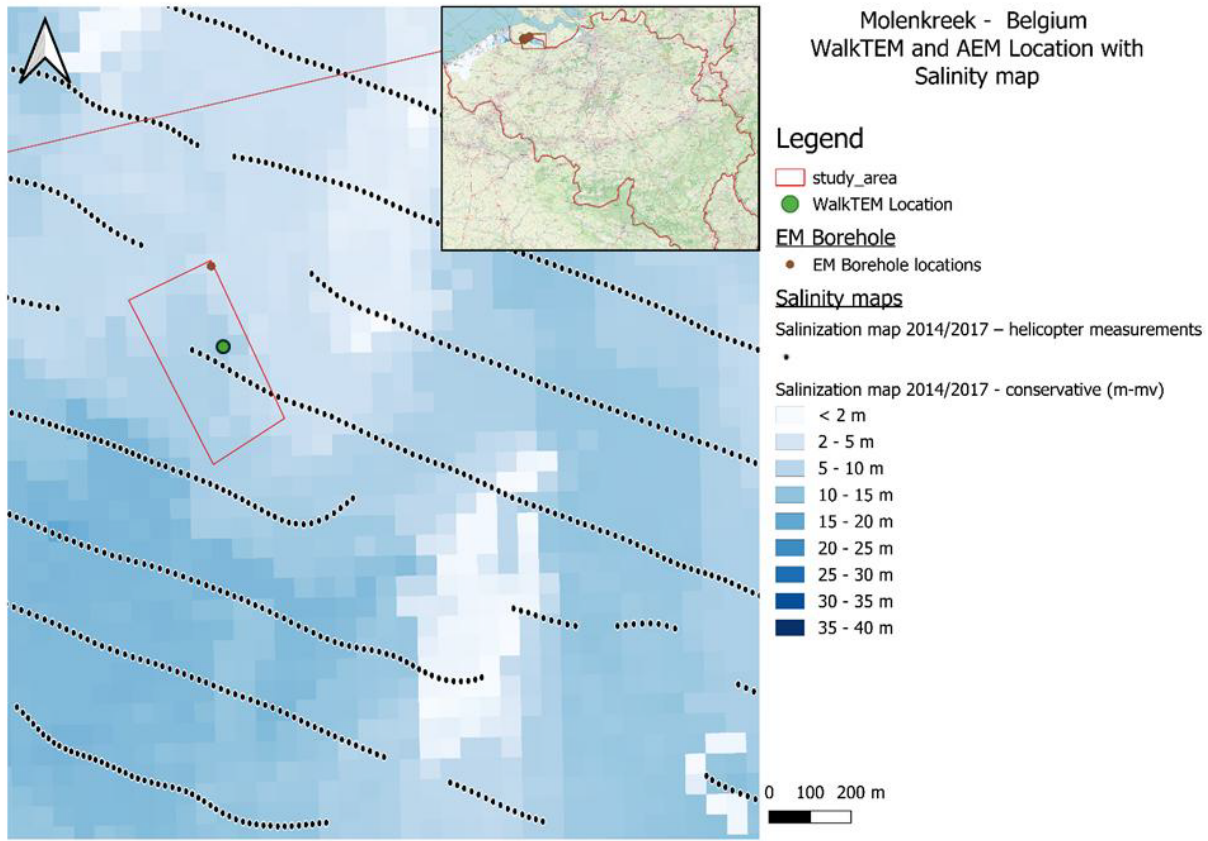

*Figure 9. Localization of the study site in the Meetjesland. The green circle shows the WalkTEM site location in the northwest of Belgium, with the black dots showing AEM soundings used to produce the FSI map shown in the background (a darker blue color means a deeper interface).*

The selection of prior distribution plays a critical role in Bayesian inference and stochastic inversion, and requires careful consideration when working with real data (Ahmed et al., 2024). The prior distributions were selected based on pre-existing geological knowledge available from public online resources in 'Databank Ondergrond Vlaanderen' and an extensive database within our own laboratory. For simplicity, uniform distributions are used for all parameters (Dramsch 2020), but the conductivity are expressed in a log-scale. The prior range for the different layers is combines information from borehole (Figure 1) with the existing salinization map. The latter includes a geological model with the depth of the underlying confining clay layer. An example of the salinity map database along the AEM flight line closest to the site is presented in Figure 10. The salinity map in the study area reveals an average 10 m thick freshwater layer, ranging between 5 m and 15 m. The transition from fresh to saline conditions extends from 5 to 15 meters below the surface, below which a saline layer exists. The clay layer is located between 17 and 30 m depth. The ranges were expanded to compensate for the smoothness constraint used in the deterministic inversion of the AEM data (Delsman et al., 2019).

The prior distribution from the salinity map was further expanded with additional information from a neighboring borehole (Figure 1B). The log shows a thin and resistive freshwater layer, the conductivity then increases progressively from 4.5 to 12 meters deep within the FSI. The conductivity shows minimal variation between 12 and 25 meters, indicating the presence of saline water. The log stops at 25 m, where the borehole reaches the underlying confining unit. Table 4 summarizes the prior distribution derived from the available information.

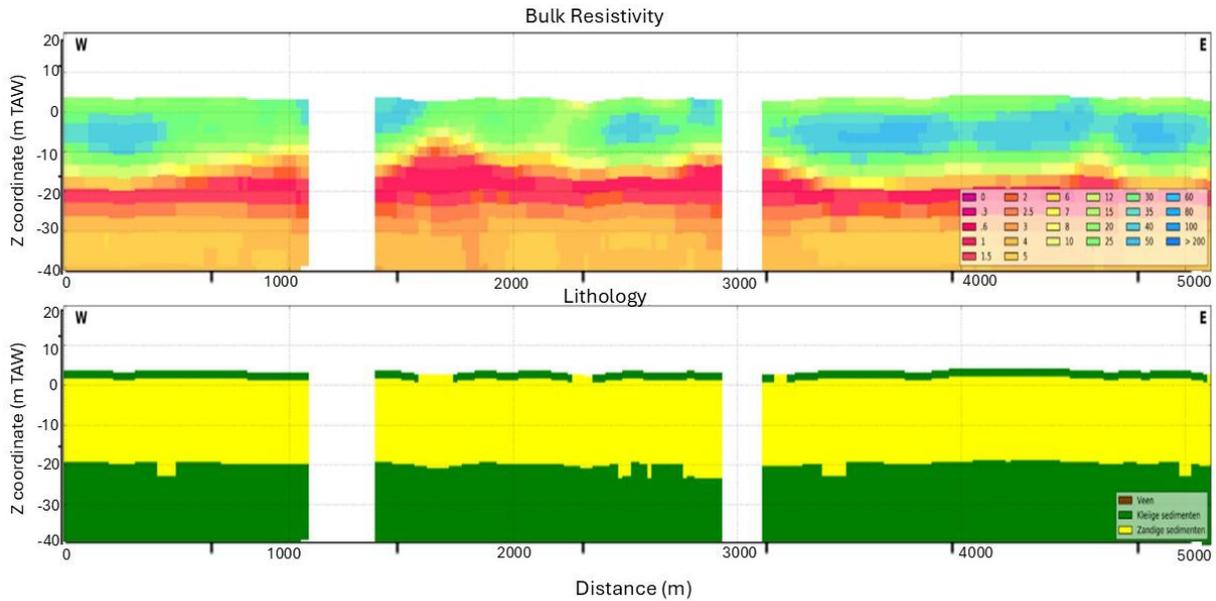

*Figure 10. Resistivity inversion of the AEM SkyTEM survey, red values correspond to low resistivity, blue values to high resistivity (top). Note the increase in resistivity in the clay layer. Geological model (yellow = sand, green = clay) corresponding to the flight line (B). Note the increase in resistivity in the clay layer (modified from Delsman et al., 2019).*

*Table 4. Prior model range for the field dataset*

| Layers | Thickness/depth (m) | | Conductivity (Sm$^{-1}$) | |
|---|---|---|---|---|
| | **Minimum** | **Maximum** | **Minimum** | **Maximum** |
| layer 1 | 1 | 15 | 0.001 | 0.15 |
| layer 2 | 0.5 | 7 | *Linearly interpolated* | |
| layer 3 | $t_3 = d_{hs} - t_1 - t_2$ | | 0.25 | 1 |
| Half-space | 17 | 30 | 0.025 | 0.5 |

The posterior model space obtained with BEL1D-T using a threshold of 0.135 is shown in Figure 11. Most parameters only have a limited reduction in uncertainty compared to the prior distribution. The first layer thickness mode is about 12 m with a range covering 10-14 m. The inversion predicts a rather sharp FSI, lower than 5 m thick, as low thicknesses dominate the posterior, and the larger prior bound is no longer represented. The depth of the salt layer is encountered at depth ranging from 13 to 18 m in most modems. The posterior model presents a restricted uncertainty for the thickness of layer 3, because of the clay depth constraint. All recovered models are located at the lower end of the prior for the clay layer depth (17.5 to 22.5 m). Although this could indicate that the prior should be extended to lower values, this is unrealistic according to prior information. In terms of conductivity, the posterior is dominated by low values within the prior range. The first layer has abimodal distributions with values ranging from 0.001 to 0.05 S/m. The estimated conductivity of the saline layer has a mode around 0.55 Sm$^{-1}$. Finally, the inversion indicates a relatively resistive value for the clay layer with a mode around 0.15 Sm$^{-1}$, high conductivities being unlikely.

To check the validity of the posterior distributions, we show the decay curve for each posterior sample. The field data falls entirely within the range of simulated data from the posterior distribution (Figure 12A) for both the high and low moments. The posterior range in the data space is relatively large at early time steps, which should display a larger signal-to-noise ratio but may also be subjected to primary-field contamination. This is likely an effect of the definition of the threshold, which is based on the rRMSE calculated globally for all time gates. Combined with a relatively conservative value for the threshold, it gives posterior samples with a larger discrepancy at the early time gates very well.

With the chosen threshold of 0.135, the rejection rate of BEL1D-T is 83% after thresholding. As an illustration, we also show the results for a threshold of 0.09, roughly corresponding to a relative error of 9% (see Equation (4) in Ahmed et al. 2024) on the data, i.e. 3 times the estimated noise level. The rejection rate increases to 92.7%. While the uncertainty is slightly reduced as the samples with a higher misfit are removed from the posterior (Figure 13), this concerns mostly outliers quite far from the mean of the distribution, so that the overall assessment of the uncertainty remains the same. In particular, the range of values for the different parameters remains similar. Nevertheless, the fit of the early time gates improves significantly (Figure 12B). This illustrates that, although it might be difficult to appropriately choose a threshold, as the actual noise level on the data is unknown, BEL1D-T provides a fast and robust approximation of the posterior distribution.

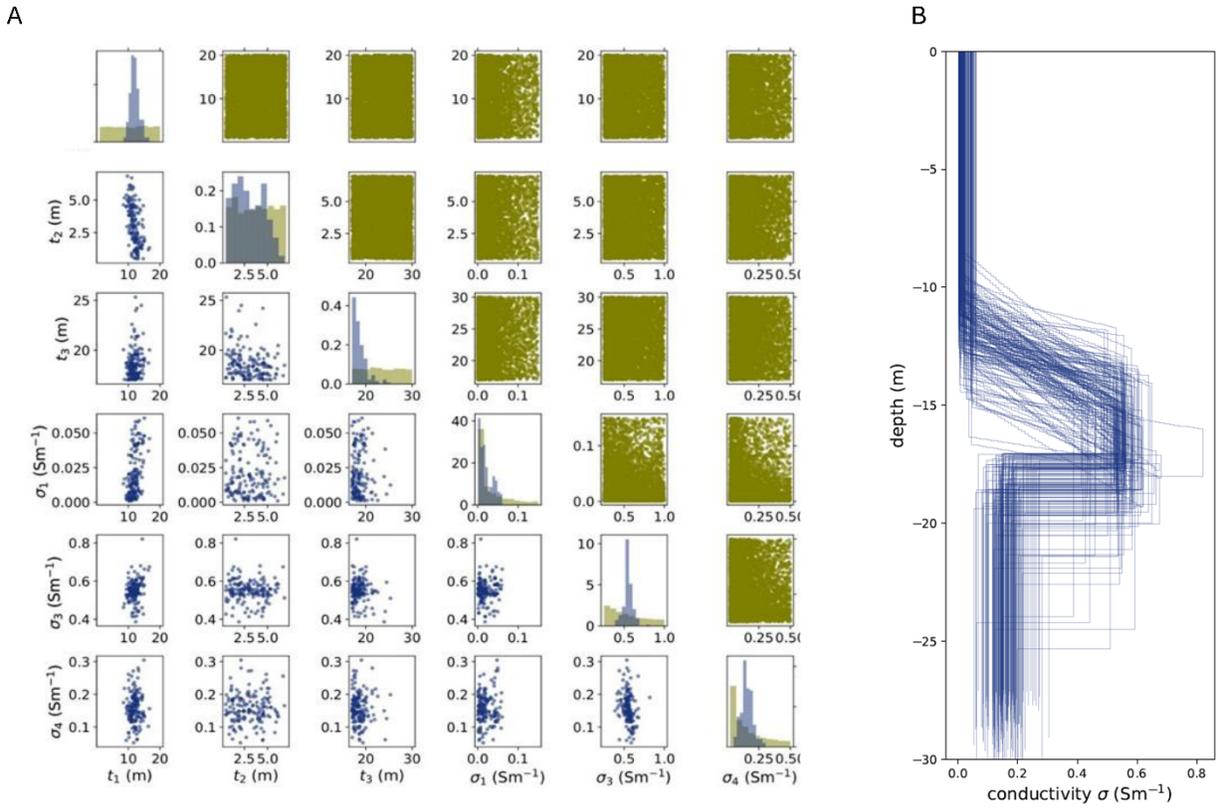

*Figure 11. Posterior model space correlations (A) and depth-conductivity posterior samples (B) for the field case with the 0.135 threshold.*

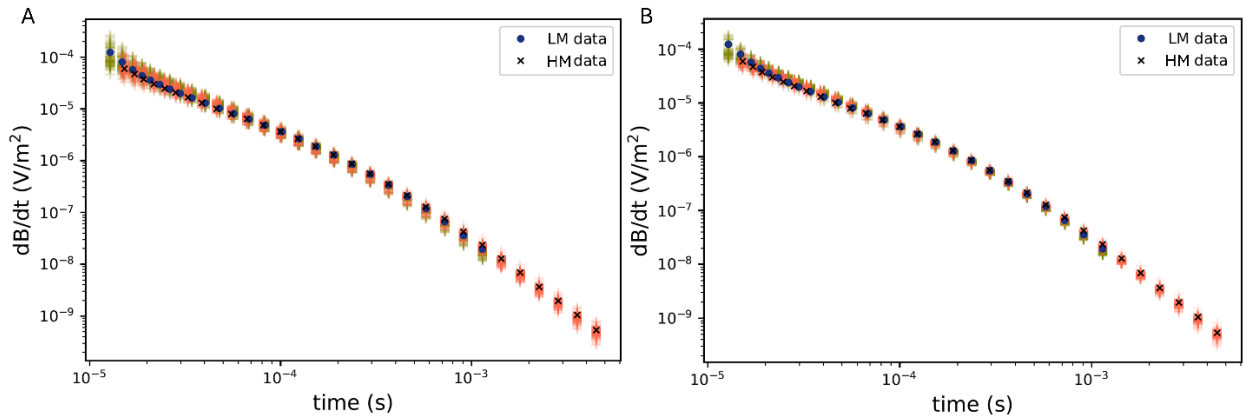

*Figure 12. Posterior data space visualization after thresholding at 0.135 (A) and 0.09 (B).*

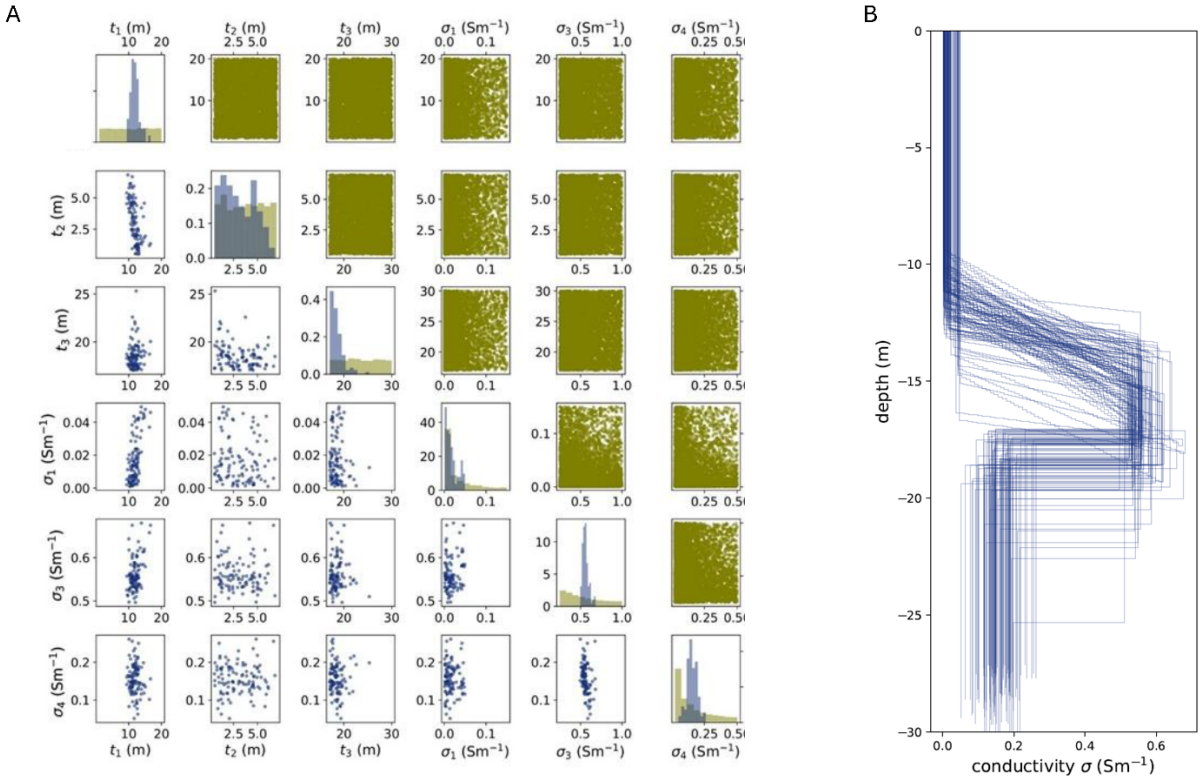

*Figure 13. Posterior model space correlations (A) and depth-conductivity posterior samples (B) for the field case with the 0.09 threshold.*

## 4. Conclusion

In this paper, we propose to combine a stochastic approach approximating the posterior distribution, namely the BEL1D-T inversion method, with a specific well-informed parameterization to invert ground-based TDEM data. The proposed parameterization incorporates the available prior knowledge on the fresh-saltwater interface with only six model parameters, significantly reducing the complexity and computational burden often associated with stochastic approaches. The parameterization represents the gradual conductivity transition generally observed in coastal settings affected by saltwater intrusion. As this transition often varies throughout the study area, the parameterization can model both sharp and gradual (linear) transitions from fresh to saltwater. Due to the explicit parameterization including the depth and the thickness of the transition zone, the uncertainty can be straightforwardly assessed. The results show that this inversion approach can effectively estimate the uncertainty range of model parameters. The BEL1D-T inversion, combined with suitable prior distributions, enabled the generation of a reliable posterior distribution to evaluate uncertainties in subsurface properties.

The results show that the TDEM data inversion successfully estimates the properties of the freshwater layer on top of the saltwater lens, but fails to significantly reduce the uncertainty on the thickness and depth of the transition zone. This lack of sensitivity of TDEM data to the transition zone is likely the cause of this limitation. In particular, the thickness of the transition zone and the properties of the shallow freshwater layer compensate for each other. The estimation of the uncertainty of the FSI seems mostly impacted by the limitations of the measuring configuration. In contrast, it is not significantly affected by the uncertainty on the conductivity of the overlying freshwater layer and underlying saline layer. These conductivities can be reliably estimated, including their uncertainty. We show that adding a constraint on the underlying confining unit, in this case a clay layer, strongly helps to retrieve the thickness of the saline layer, but it only slightly impact the posterior distribution of the fresh-saline interface.

For the field case, the existing salinity maps, the geological model, and local EM logs allowed for the definition of realistic prior ranges for the model parameters, resulting in turn in reliable posterior distribution estimates. Applying the clay depth constraint based on known aquifer thickness ranges strongly helped to reduce uncertainty on the saline layer thickness. It is a parameter that can often be estimated using existing data and should therefore be included in the parameterization and the prior space.

The proposed framework currently estimates parameter uncertainty based on a chosen parameterization. Future work could incorporate hierarchical Bayesian models or formal model selection techniques to explicitly address structural uncertainty. This could, for example, allow for the inversion of TDEM data without knowing in advance if an FSI is present, add more transition or more layers. In addition, the role of the 1D forward model and possible related modeling errors in the inversion process could be addressed. Next to multi-dimensional modelling error, our parameterization also ignores within layer variations that could affect the forward response. Finally, other EM acquisition designs, both in the time-domain and frequency domain, could be examined to identify configurations more sensitive to the fresh-saltwater interface.